\documentclass{optica-article}

\journal{opticajournal} 

\articletype{Research Article}

\usepackage[utf8]{inputenc}
\usepackage{subcaption} 
\usepackage{xcolor}
\begin{document}

\title{Efficient Mode Conversion at 1064 nm via Bilayer Inverse Taper on Thin-Film Lithium Niobate}

\author{Ruidong Xue,\authormark{1,*} Jobayer Hossain,\authormark{2} Joshua Arnold,\authormark{1} Jiuyi Zhang,\authormark{1} Xiaofeng Zhu,\authormark{1} Marco Moller de Freitas,\authormark{3} Christopher J Cullen,\authormark{1} Shouyuan Shi,\authormark{1,3} Peng Yao,\authormark{1} Timothy Creazzo,\authormark{1} and Dennis W. Prather\authormark{1,3}}

\address{
\authormark{1}Phase Sensitive Innovations, Inc., Newark, DE 19713, USA\\
\authormark{2}School of Electrical Engineering \& Computer Science, University of North Dakota, Grand Forks, ND 58202, USA\\
\authormark{3}Department of Electrical and Computer Engineering, University of Delaware, Newark, DE 19716, USA
}

\email{\authormark{*}xue@phasesensitiveinc.com} 


\begin{abstract*} 

Efficient fiber-to-chip coupling remains challenging for thin-film lithium niobate (TFLN) devices operating at 1064 nm. Here, we fabricate and characterize an on-chip optical mode converter based on a bilayer inverse taper. The proposed structure demonstrated a coupling loss of 1.9 dB per facet for the TE mode, with a 1 dB bandwidth spanning 1055--1085 nm. The polarization-dependent alignment tolerance was measured along the lateral and vertical directions. Theoretical and simulation analyses indicate that, when using a lensed fiber with a mode field diameter of 2.5 $\mu$m, the coupling loss can be reduced to as low as 0.48 dB per facet. These results demonstrate the feasibility of providing a high-efficiency solution for on-chip optical coupling at 1064 nm.

\end{abstract*}

\section{Introduction}
Unlike electrons that traverse copper interconnects, photons are free from RC delay effects associated with parasite wiring capacitance. As a result, photonic interconnects can deliver extremely high operational speeds together with ultra-low power consumption in modulation\cite{daudlin2025three,zhou2024silicon}. Among various platforms, silicon photonics has developed most rapidly due to its compatibility with CMOS processes, which has led to high-speed PIN modulators, including ring resonator structures\cite{yuan20245,chan2022c,chan2023efficient} and Mach-Zehnder Modulator (MZM) structures\cite{li2023integrated}, as well as Ge-based heteroepitaxial photodetectors\cite{lischke2021ultra,virot2014germanium}. However, when attempting to further improve the upper speed limit of silicon-based photonic devices, inherent material properties impose fundamental physical constraints. Thus, driven by the growing demands of AI computation and big data transmission, there is an urgent need for physical platforms that can support even higher modulation speeds. In comparison, thin-film lithium niobate (TFLN) has emerged as a promising candidate for next-generation high-speed on-chip modulators and quantum sensors, owing to its unique nonlinear properties\cite{zhang2023quantum}. Today, TFLN modulators can approach terahertz operation while achieving on-chip propagation losses below 0.2 dB/m\cite{lampert2025photonics,shams2022reduced,zhu2025capacitive}. Despite challenges such as the difficulty of etching TFLN during device fabrication and the need to ensure impedance matching between RF transmission lines and optical waveguides, to minimize bandwidth degradation caused by high-frequency mismatches, TFLN still demonstrates exceptional performance. It enables extremely low V$_\pi$ and ultra-low loss features that are difficult to realize in traditional silicon photonic platforms. In addition, reducing the operating wavelength provides a direct route to lowering the half-wave voltage \(V_{\pi}\) of TFLN electro-optic modulators. For a Pockels-effect phase modulator, the half-wave voltage scales approximately as \(V_{\pi}\propto \lambda/(n^3\gamma_{\mathrm{eff}}\Gamma)\), where \(\lambda\) is the operating wavelength, \(n\) is the optical refractive index, \(\gamma_{\mathrm{eff}}\) is the effective electro-optic coefficient, and \(\Gamma\) is the optical-electrical overlap factor. This relation shows that reducing the wavelength from 1550 nm to 1064 nm can directly lower \(V_{\pi}\), while changes in optical confinement may further reduce \(V_{\pi}\) indirectly through the geometry-dependent overlap factor \(\Gamma\)\cite{renaud2023sub}.

However, fiber-to-chip coupling remains an unavoidable challenge. Thus, to efficiently couple light from optical fiber into photonic chips, two main structures are typically employed. The first is the Bragg grating coupler, which utilizes phase matching in reciprocal space to couple free-space light into the chip at specific angles, achieving very low insertion loss\cite{vitali2023high,sanchez2024ultra,liao2024high,wang2017unidirectional}. However, this structure suffers from limited bandwidth and sensitivity to the incident angle of the fiber. The second structure is the waveguide inverse taper coupler, where the fundamental mode of the optical fiber is coupled to a  waveguide with an expanded optical mode at the edge of the chip\cite{he2019low,hu2021high,liu2022ultra,liang2022efficient,chen2023low,zhang2023efficient,jia2023high,guo2024polarization,chen2024high}. Through adiabatic transition, the fundamental mode of the waveguide is transformed into a mode that more closely matches the mode of the optical fiber. In the C-band, both silicon photonics and lithium niobate platforms have achieved very low insertion losses using this approach. Beyond the C-band, however, there is growing demand for low-insertion-loss coupling at other wavelength ranges. At 1310 nm and in the visible spectrum, coupling losses as low as 1 dB and 3 dB per facet separately have already been reported\cite{guo2024polarization,liu2022ultra}. Recently, Zhang et al. reported a 1064-nm TFLN electro-optic modulator incorporating a double-layer tapered edge coupler designed for a single-mode fiber of 5.3 \(\mu\) m mode field diameter, with a measured end-face coupling loss of approximately 3 dB per facet \cite{zhang2026cltwe}. However, efficient TFLN edge coupling at 1064 nm has not been extensively investigated, despite the importance of this wavelength for emerging TFLN modulators and RF-photonic systems that benefit from the shorter-wavelength scaling of $V_\pi$\cite{zhang2025high,jagatpal2021thin,de2026thin}. Existing bilayer inverse tapers and nonlinear taper profiles provide valuable design concepts, but they cannot be directly transferred to the 1064 nm regime because the reduced wavelength changes the modal confinement, fiber-to-chip mode mismatch and adiabatic mode conversion condition. These effects require a wavelength-specific design and experimental investigation. The opportunity to build RF photonic systems at 1064 nm to leverage lower $V_\pi$ devices gives rise to an urgent need to develop an efficient on-chip coupler for 1064 nm. For this reason, this work focuses on the design, fabrication, and testing of high-efficiency on-chip couplers based on TFLN. To this end, simulations were performed to study inverse taper device parameters, including taper length and taper adiabatic behavior, and evaluate their impact on coupling loss \cite{xue2026spie}. The analysis shows that lensed fibers can achieve coupling losses as low as 0.48 dB per facet. The fabricated devices demonstrate a loss of 1.9 dB per facet for the TE mode, with a 1 dB bandwidth spanning 1055--1085 nm. The proposal and realization of an on-chip coupler at 1064 nm paves the way for future integration of on-chip devices operating in this wavelength band.

\section{Design and Simulation}
\subsection{Dual layer inverse taper design}
\begin{figure}[htbp]
    \centering

    \begin{subfigure}{0.95\textwidth}
        \centering
        \includegraphics[width=\textwidth]{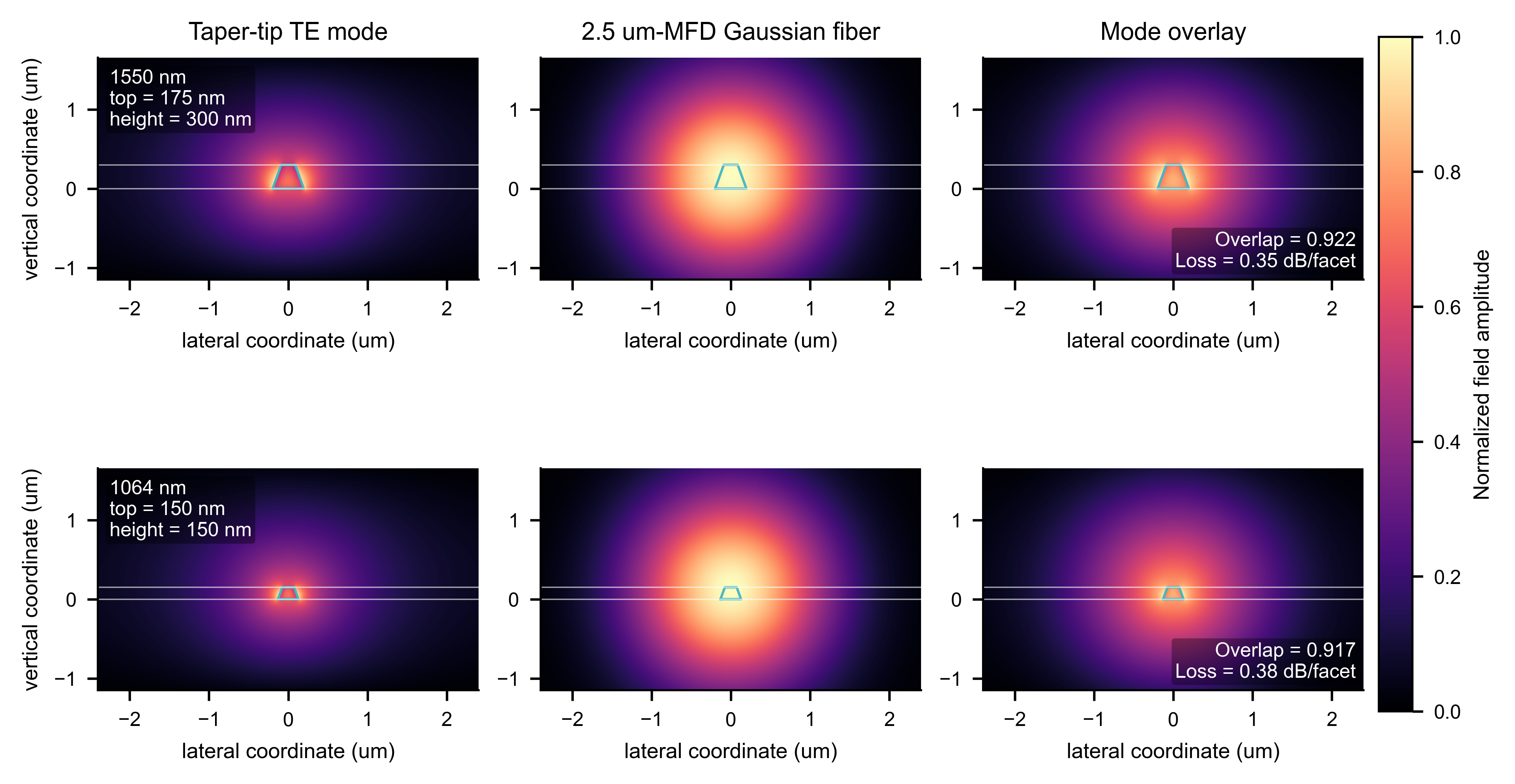}
        \caption{}
        \label{fig:fig1a}
    \end{subfigure}
    \hfill
    \begin{subfigure}{0.45\textwidth}
        \centering
        \includegraphics[width=\textwidth]{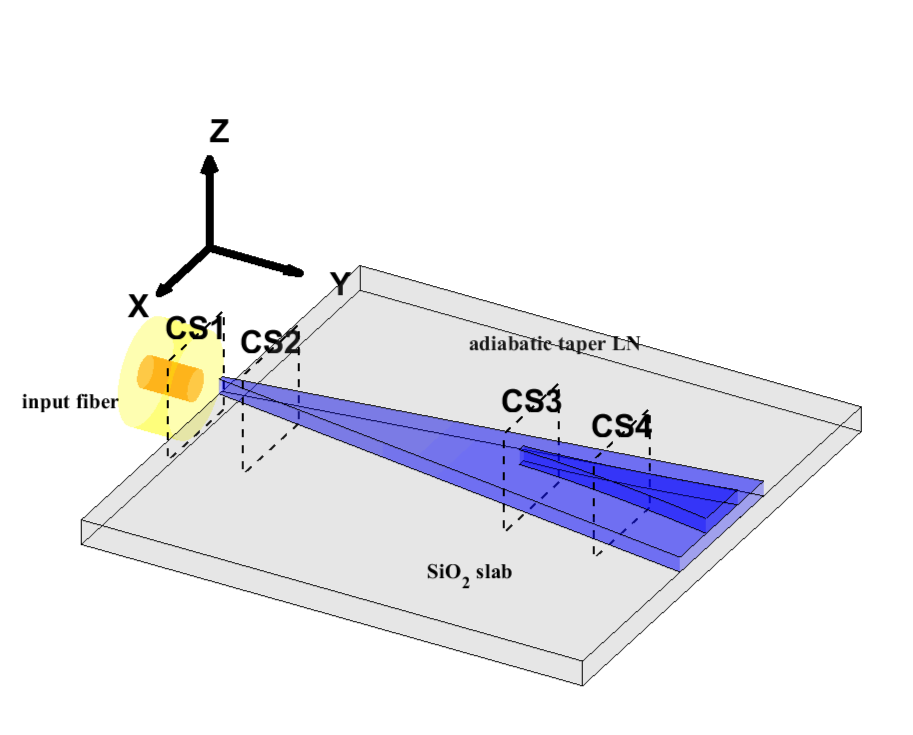}
        \caption{}
        \label{fig:fig1b}
    \end{subfigure}

\vspace{1ex} 

    \begin{subfigure}{0.22\textwidth}
        \centering
        \includegraphics[width=\textwidth]{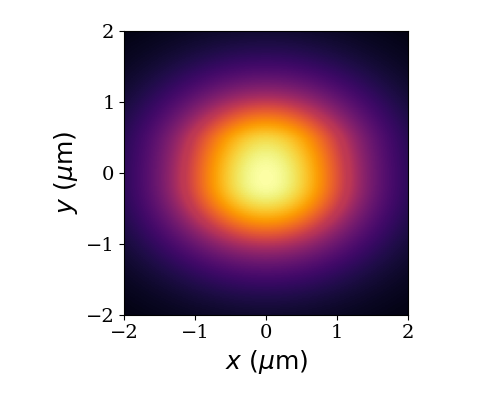}
        \caption{}
        \label{fig:fig1c}
    \end{subfigure}
    \hfill
    \begin{subfigure}{0.22\textwidth}
        \centering
        \includegraphics[width=\textwidth]{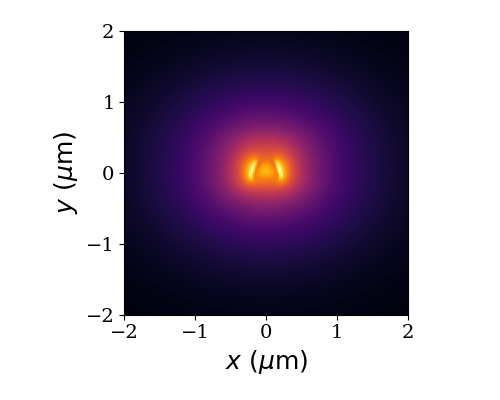}
        \caption{}
        \label{fig:fig1d}
    \end{subfigure}
    \hfill
    \begin{subfigure}{0.22\textwidth}
        \centering
        \includegraphics[width=\textwidth]{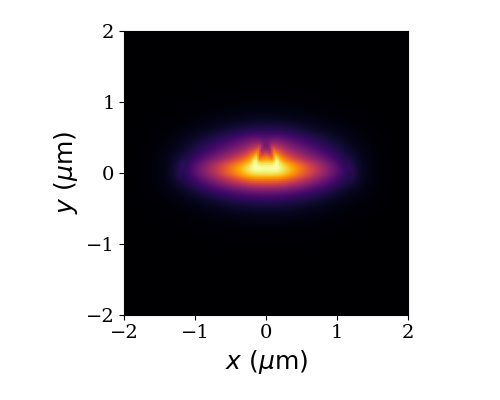}
        \caption{}
        \label{fig:fig1e}
    \end{subfigure}
    \hfill
    \begin{subfigure}{0.22\textwidth}
        \centering
        \includegraphics[width=\textwidth]{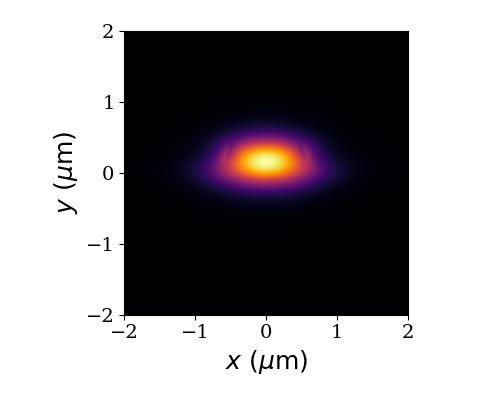}
        \caption{}
        \label{fig:fig1f}
    \end{subfigure}

    \caption{(a) Wavelength specific fiber to taper mode overlap comparison for the 1550 nm reference taper and the optimized 1064 nm taper using a 2.5~$\mu$m MFD Gaussian fiber mode.
    (b) overall 3D schematic of the bilayer inverse taper, and (c--f) simulated TE mode profiles in waveguide cross-sections CS1--CS4.}
    \label{fig:illustration_fig1}
\end{figure}

In the design of an inverse taper, reducing optical loss is the most important challenge. 
The main source of loss is the mode mismatch between the single-mode fiber and the input facet. Typically, the fiber mode is much larger than the on-chip optical waveguide mode resulting in significant loss when coupling between the two. In a standard single-layer inverse taper, the waveguide width is reduced squeezing the optical mode outwards laterally. In many cases, this lateral expansion of the mode is not sufficient for achieving the lowest possible optical coupling loss. As a result, researchers began using bilayer tapers that facilitate transition of the optical waveguide mode into a thinner final layer that is inverse tapered, which facilitates expansion of the mode in the lateral and vertical directions. At 1550 nm, a 600 nm ridge etched waveguide with a 300 nm etch depth is typically used\cite{jin2021efficient,li2024linearity} resulting in a final taper layer thickness of 300 nm. However, the 1064 nm coupler cannot be treated as a simple scaled version of a bilayer taper designed at 1550 nm. The TFLN film thickness, taper layer thickness, input mode size, and taper transition profile must be re-optimized accordingly to reach the optimum point of mode matching and adiabatic conversion. To identify a suitable 1064 nm taper-input geometry, we used the effective index comparison only as a qualitative design reference rather than as a direct measure of modal confinement. The reduced wavelength changes the modal field distribution and fiber to chip mode mismatch, so the final design should be evaluated by the overlap between the incident fiber mode and the taper input mode. As shown in Fig.~1a, we therefore performed a wavelength specific fiber to taper mode overlap comparison using a 2.5~$\mu$m mode field diameter (MFD) Gaussian fiber mode. For the 1550 nm reference taper input with a 175 nm top width and a 300 nm input height, the calculated overlap is 0.922, corresponding to a coupling loss of 0.35~dB per facet. For the optimized 1064 nm taper input with a 150 nm top width and a 150 nm input height, the calculated overlap is 0.917, corresponding to a coupling loss of 0.38~dB per facet. This mode overlap comparison shows that the 1064 nm taper geometry was optimized to maximize fiber to taper mode matching comparable to the 1550 nm reference design. In our design, we select a 300 nm thick TFLN waveguide with a 150 nm etch depth. As shown in Fig~\ref{fig:fig1b}, the bi-layer inverse taper structure couples the Gaussian beam into the ridge-etched TFLN waveguide. The input waveguide port has a width of 150 nm and a height of 150 nm, while the simulated fiber fundamental mode has a 2.5~$\mu$m Gaussian mode-field diameter. The TFLN ridge-etched waveguide has an 800 nm width, 300 nm thickness, and 150 nm etch depth to ensure only fundamental modes are allowed. In the mode and coupling simulations, the waveguide cross-sections were modeled as ridge waveguides with non-vertical etched sidewalls. The cross-sectional dimensions described above therefore refer to the nominal top width, total TFLN thickness, and etch depth used in the device layout.

We simulated the waveguide modes of four different cross sections (see Fig~\ref{fig:fig1c}--\ref{fig:fig1f}) along the taper. At the input port, the optical mode is expanded to match the fiber mode, and as the propagation distance increases, the TE mode gradually couples from the first layer to the second layer, eventually transforming into the fundamental TE mode of the ridge waveguide. During this gradual conversion process in the inverse taper, some of the optical energy is lost due to the nonadiabatic mode transition caused by rapid geometric changes. Thus, the design of a high-efficiency coupler requires both mode matching at the input port and adiabatic geometric transition of the taper. In previous studies, most of the design has adapted linearly varying taper lengths, indicating a long geometrical length is required to maintain adiabatic mode conversion. In this work, a nonlinear tapered design at 1064 nm is proposed to reduce propagation loss, inspired by previous studies that employed nonlinear tapers for mode conversion\cite{liu2024high,kaushalram2020mode,hosseini2010mode}. The geometric variation of the second-layer waveguide is described by the following power-law nonlinear equation:

\begin{equation}
y(x) = w_r + (w_l - w_r)\left(1 - \frac{x}{L}\right)^m, \quad 0 \le x \le L,
\end{equation}
where $w_r$ and $w_l$ denote the right and left taper widths, respectively, $y(x)$ represents the width of the taper at position $x$, and $m$ is the exponent parameter that controls the rate of taper variation. It can be observed that when $x = 0$ and $x = L$, the function corresponds to the left and right ends of the taper, respectively. When $m = 1$, the equation describes a linear taper. For $m<1$, the width variation is slowed in the initial section of the lower taper. This profile is useful because the optical mode is weakly confined near the taper input and is highly sensitive to abrupt geometric changes. A slower initial transition therefore helps suppress nonadiabatic coupling while keeping the total taper length compact.

\subsection{Device Simulation and Optimization}
\begin{figure}[htbp]
    \centering

    \begin{subfigure}{0.45\textwidth}
        \centering
        \includegraphics[width=\textwidth]{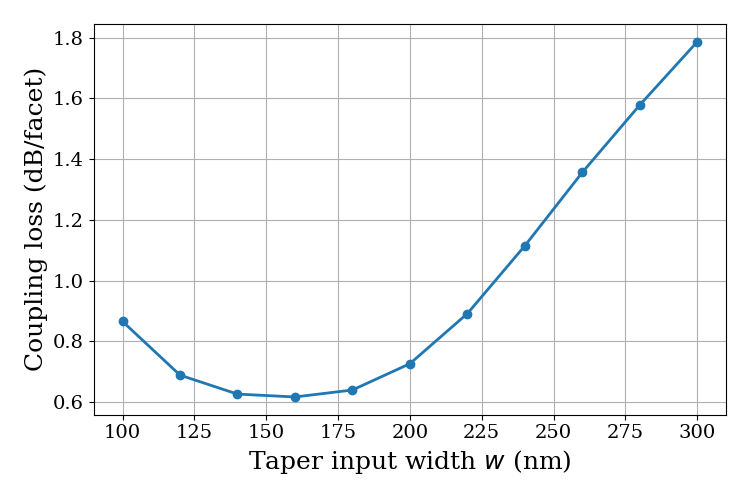}
        \caption{}
        \label{fig:fig2a}
    \end{subfigure}
    \hfill
    \begin{subfigure}{0.45\textwidth}
        \centering
        \includegraphics[width=\textwidth]{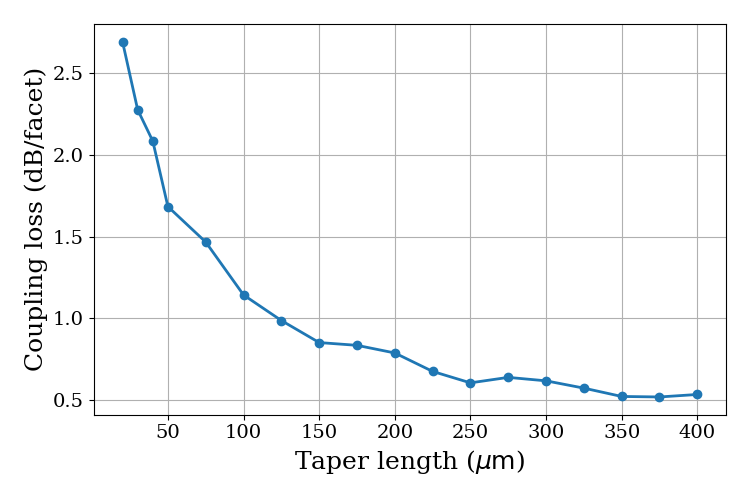}
        \caption{}
        \label{fig:fig2b}
    \end{subfigure}

    \vspace{1ex} 

    \begin{subfigure}{0.45\textwidth}
        \centering
        \includegraphics[width=\textwidth]{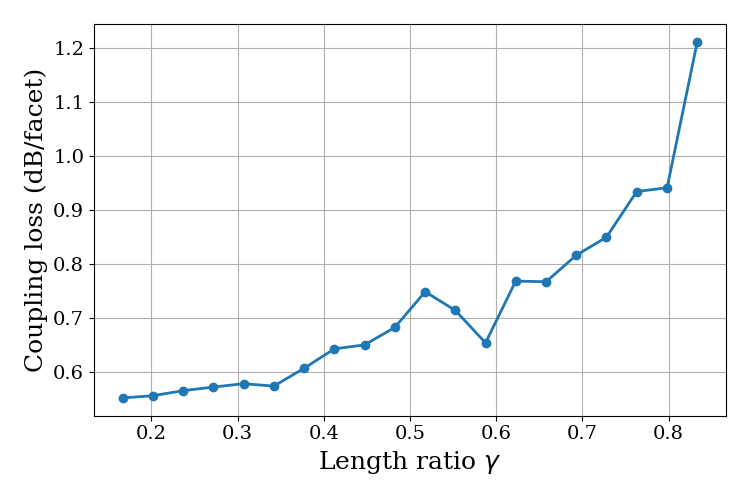}
        \caption{}
        \label{fig:fig2c}
    \end{subfigure}
    \hfill
    \begin{subfigure}{0.45\textwidth}
        \centering
        \includegraphics[width=\textwidth]{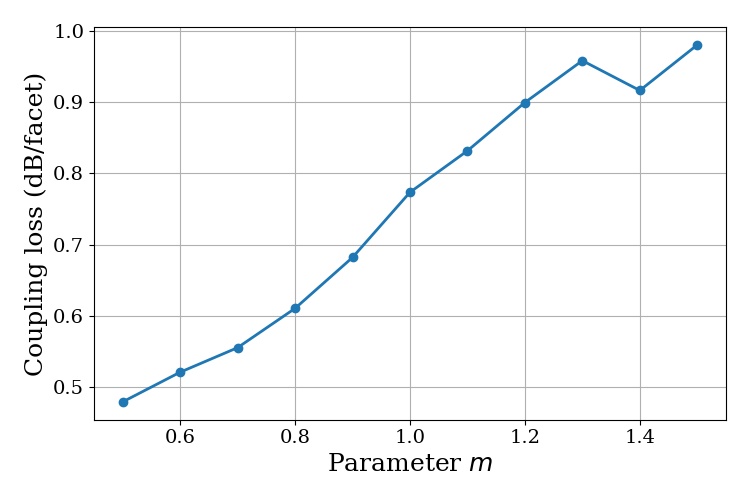}
        \caption{}
        \label{fig:fig2d}
    \end{subfigure}

    \caption{Parameter sweep results for the adiabatic taper geometry design: (a) simulated coupling loss as a function of taper input width $w$, (b) simulated coupling loss as a function of taper length, (c) simulated coupling loss as a function of taper length ratio $\gamma$, and (d) simulated coupling loss as a function of exponent $m$.}
    \label{fig:transmission_taper_array}
\end{figure}

To optimize the design of the inverse taper, a systematic investigation of the influence of four parameters was performed : input width, $w$, length ratio, $\gamma$, and exponent, $m$ and total taper length on the taper coupling loss based on the aforementioned equations. The length ratio represents the relative starting position of the second-layer taper with respect to the first layer length. Using the Lumerical FDTD mode solver, the total coupling loss from the fiber to the ridge-etched waveguide was simulated, as shown in Fig.~\ref{fig:fig2a}. In the simulation, the second-layer fiber coupling facet used a fundamental fiber mode with a 2.5~$\mu$m mode field diameter (MFD), with a 30 nm input taper width for the second layer, an 800 nm waveguide width, and a 150 nm etch depth. The effect of the input waveguide width on coupling loss was analyzed, with the input width scanned from 100 nm to 300 nm, as shown in Fig.~\ref{fig:fig2a}. The minimum coupling loss of 0.6~dB per facet was obtained at approximately 160 nm, while widths either larger or smaller than this value led to increased mode mismatch and additional coupling loss at the facet. As shown in Fig.~\ref{fig:fig2b}, the parametric sweep of the bilayer inverse taper length shows a significant coupling loss when the taper length is below 250 $\mu $m. As the length further increases, the coupling loss stabilizes at 0.55 dB per facet. When the second-layer taper length ratio varies from 0 to 0.4, the coupling loss remains nearly constant, but it increases rapidly beyond this range. Therefore, a length ratio of 0.3 was chosen as the optimal starting point for the second-layer taper, as shown in Fig.~\ref{fig:fig2c}. Finally, the influence of the geometric shape factor $m$ of the second-layer taper on the coupling performance was studied, as shown in Fig.~\ref{fig:fig2d}. When the taper length is fixed at 300 $\mu m$, the coupling loss increases significantly when $m>0.5$. The exponent $m$ controls the taper rate of the power-law nonlinear profile. For $m=1$, the taper becomes linear. For $m<1$, including the optimized value $m=0.5$, the profile slows the geometric variation in the initial section of the lower taper, where the mode is weakly confined and particularly sensitive to geometric variations. This gradual initial transition reduces nonadiabatic coupling compared with a linear profile for the same 300~$\mu$m taper length. For larger values of $m$, the taper transition becomes less adiabatic and the simulated coupling loss increases. Therefore, $m=0.5$, with a corresponding simulated coupling loss of 0.48 dB per facet, was selected for the layout design. These results indicate that the proposed structure reveals good potential candidate for use in optical packaging applications. 

\section{Device fabrication}

\begin{figure}[htbp]
\centering

\begin{subfigure}{0.6\textwidth}
    \centering
    \includegraphics[width=\textwidth]{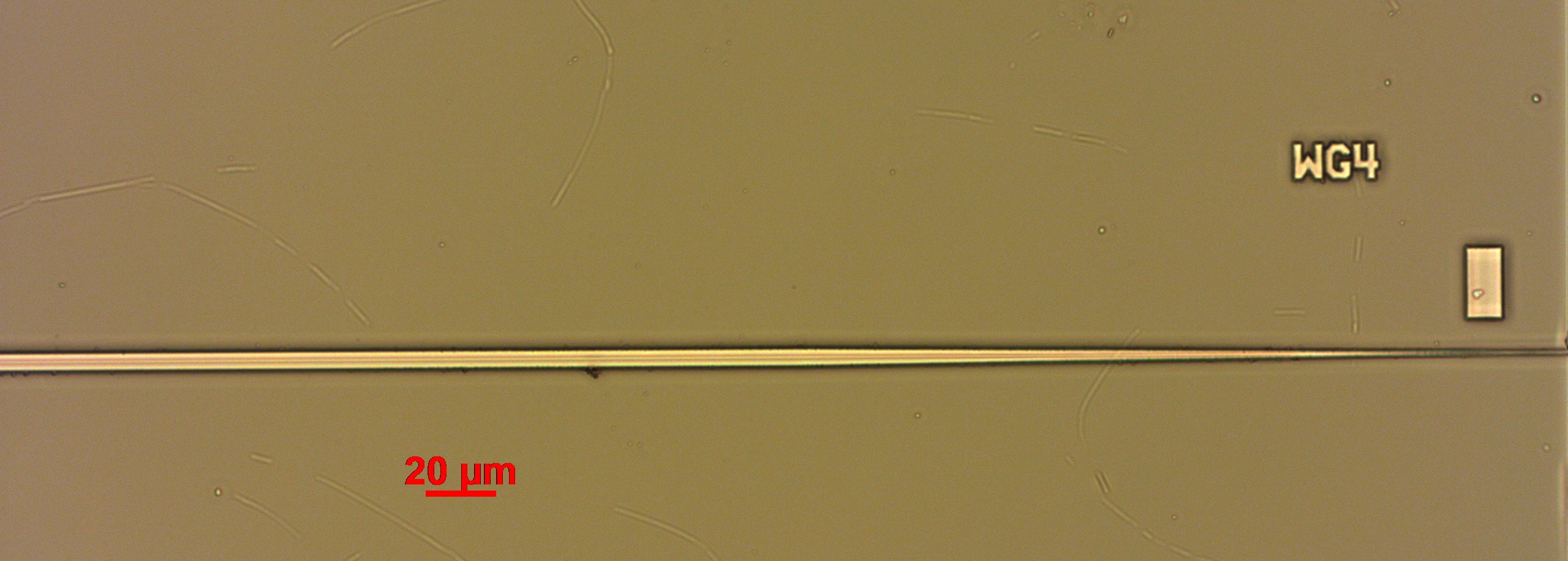}
    \caption{}
    \label{fig:optical}
    \hfill
\end{subfigure}

\vspace{1ex}

\begin{subfigure}{0.3\textwidth}
    \centering
    \includegraphics[width=\textwidth]{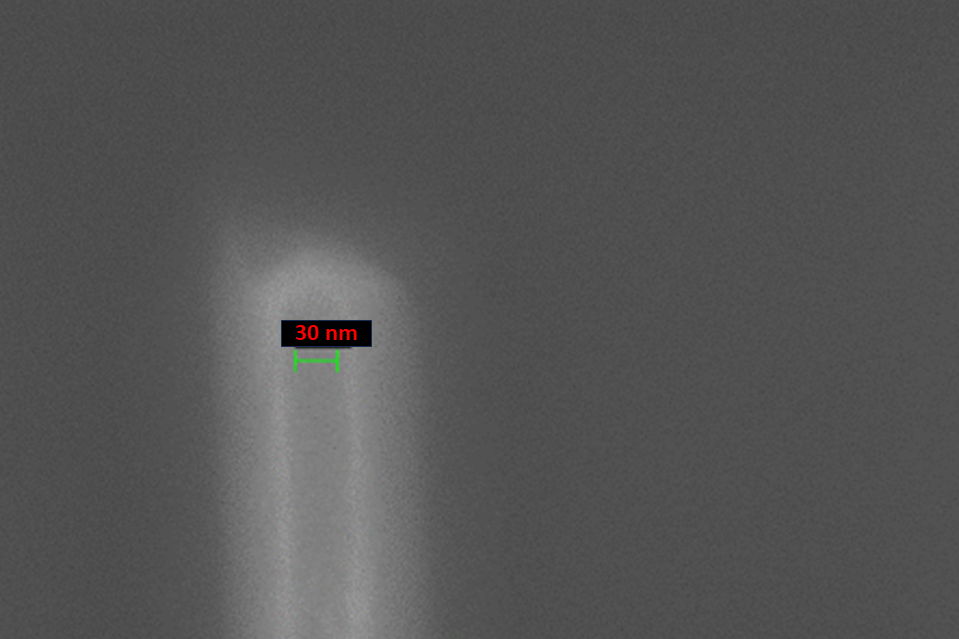}
    \caption{}
    \label{fig:SEM_a}
\end{subfigure}
\hfill
\begin{subfigure}{0.3\textwidth}
    \centering
    \includegraphics[width=\textwidth]{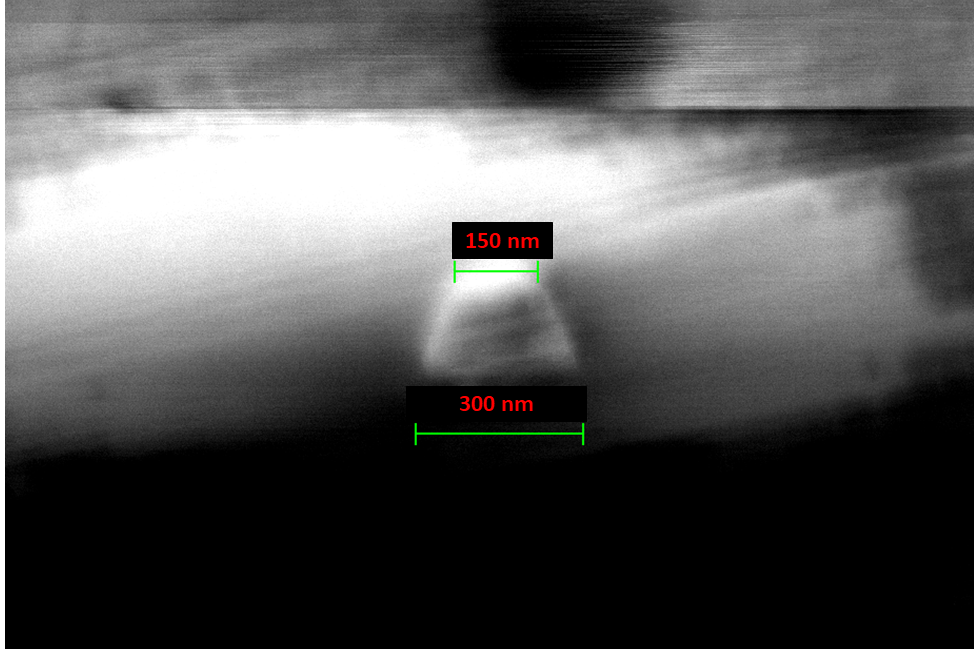}
    \caption{}
    \label{fig:SEM_b}
\end{subfigure}
\hfill
\begin{subfigure}{0.3\textwidth}
    \centering
    \includegraphics[width=\textwidth]{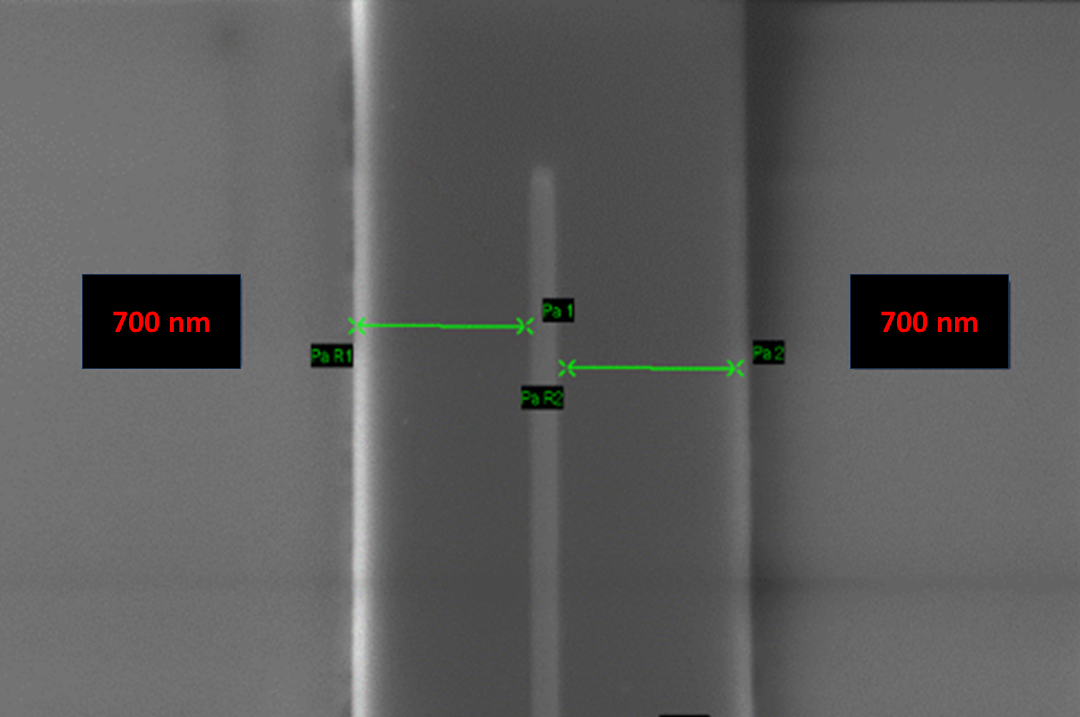}
    \caption{}
    \label{fig:SEM_c}
\end{subfigure}
\caption{(a) Optical microscope image of the fabricated device. (b) Second-layer pattern showing the finest feature size of 30 nm. (c) Cross-sectional SEM image of the fabricated waveguide. (d) Device alignment configuration.}
\label{fig:SEM_optical}
\end{figure}

The device was fabricated from a 300-nm-thick TFLN 4-inch wafer, which was diced into 1 inch pieces. Alignment marks were patterned in a 50 nm gold layer deposited via electron beam evaporation. The waveguides and tapers were patterned using  HSQ, which was spin-coated to a thickness greater than 600 nm to ensure the mask did not degrade during the waveguide etch process. High resolution electron beam lithography with 2 nA beam current was employed to define the small features of the taper, and 100 nA current was employed to define the large features including an alignment marker shielding area and facet polishing markers. AZ MIF 300 developer containing 2.5 $\%$ TMAH was used to develop the HSQ patterns after exposure. A post-develop bake with a 90°C hot plate was used to gently evaporate residual liquid, preventing structural damage and improving the selectivity of the following etching step. SEM inspection was performed after development to confirm the resist was completely removed. An Ar-ion milling etch process was carried out to define the waveguide, followed by an RCA-1 clean at 50°C to remove re-deposition on the surface and a BOE clean to remove the HSQ mask. As shown in the Fig.~\ref{fig:SEM_a} and Fig.~\ref{fig:SEM_c}, a feature size of 30 nm was achieved at the tip of the taper, and smooth waveguide sidewalls were observed under SEM.  The second layer of the bilayer inverse taper was also patterned using electron beam lithography with critical alignment to the first layer. The second layer followed a similar process to the first including an HSQ mask, Ar-ion milling etch, and post-etch wet chemical treatment to remove re-deposition. After the second layer was etched and chemically cleaned, the device alignment was inspected via SEM, which was less than 40 nm, as shown in Fig.~\ref{fig:SEM_c}. The device was clad with a 1.5~$\mu$m-thick SiO$_2$ layer deposited by plasma-enhanced chemical vapor deposition (PECVD). The chip was then diced and polished to expose the taper input facet. The polishing process included a manual polishing stage to allow for precise control over the end point of the polish, which was required to be at the tip of the second layer taper. An optical microscope image of the fabricated device is shown in Fig.~\ref{fig:optical}.

\section{Testing Setup and Results}
\begin{figure}[htbp]
    \centering
    \includegraphics[width=0.6\textwidth]{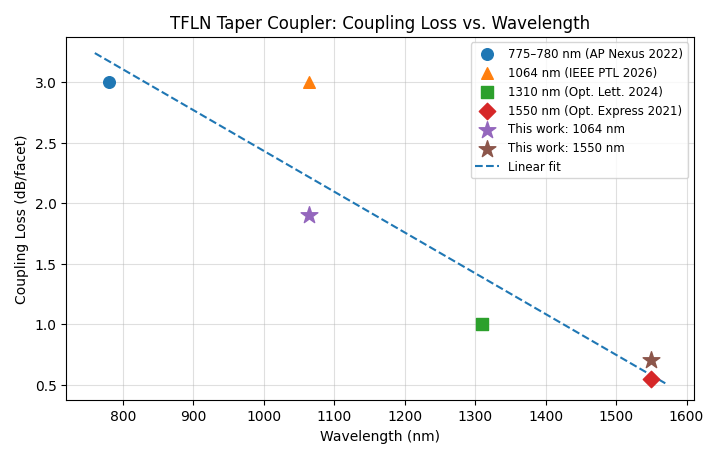}
    \caption{Measured coupling loss as a function of wavelength.}
    \label{fig:wavelength_coupling}
\end{figure}

\begin{figure}[htbp]
    \centering
    
    \begin{subfigure}{0.85\textwidth}
        \centering
        \includegraphics[width=\textwidth]{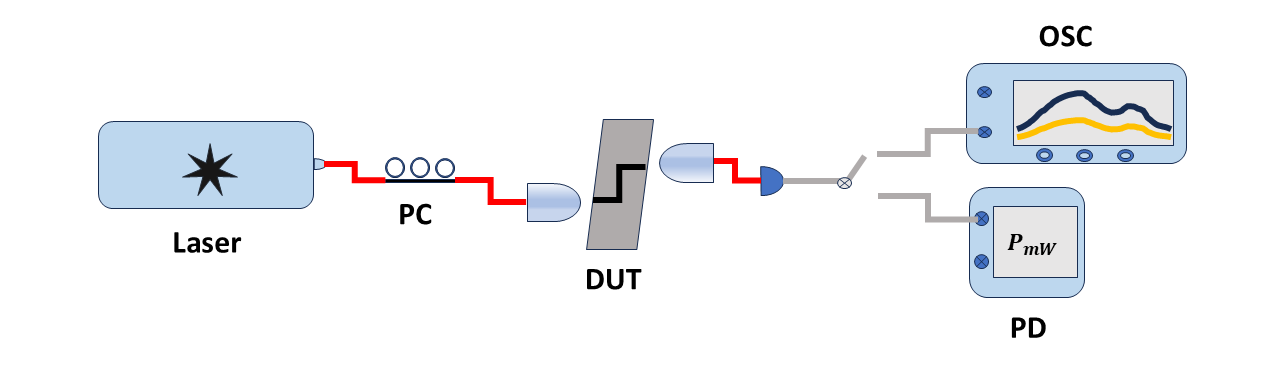}
        \caption{}
        \label{fig:setup_a}
    \end{subfigure}
    
    \vspace{1em} 
    
   \begin{subfigure}{0.28\textwidth}
    \centering
    \includegraphics[width=\textwidth]{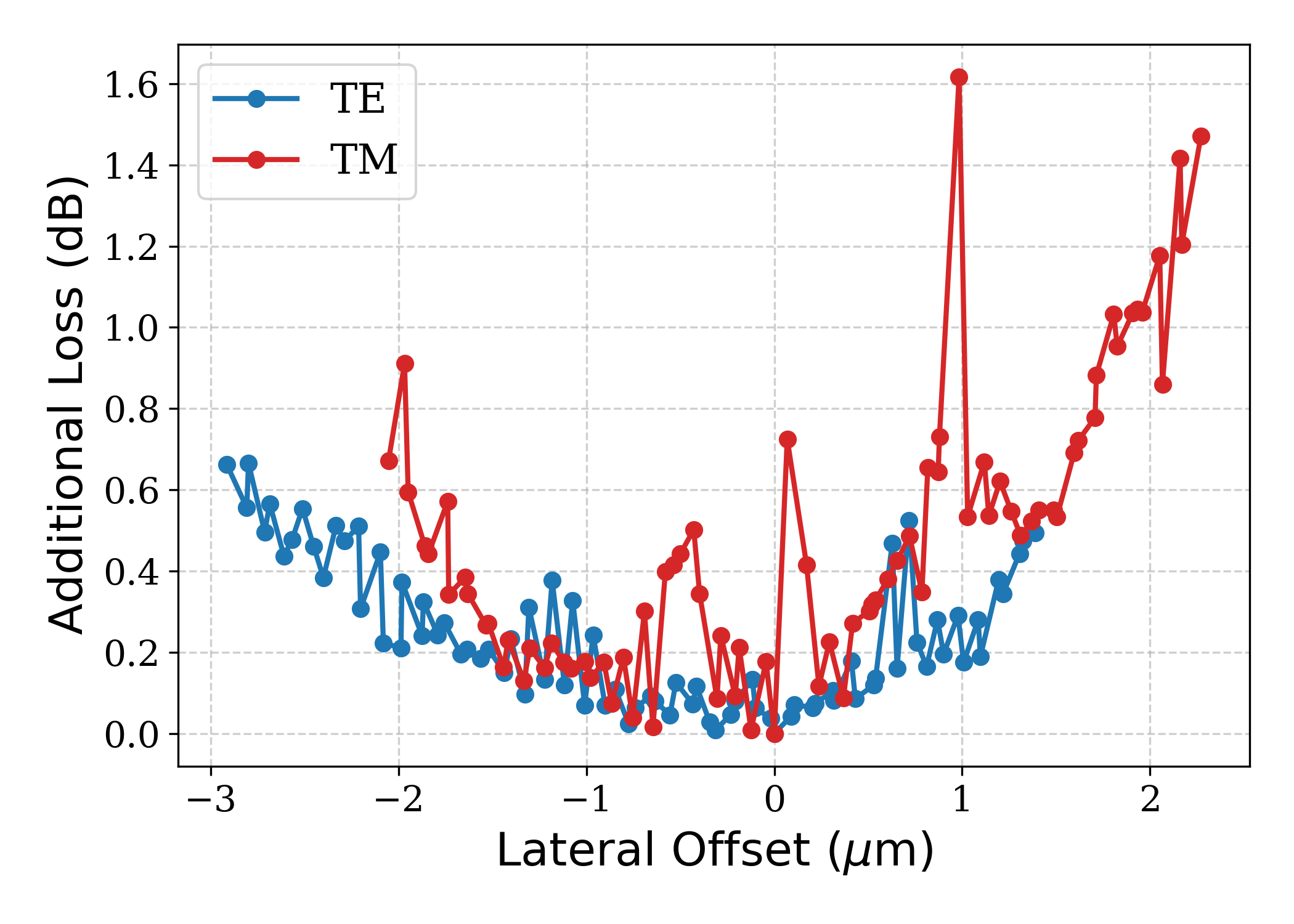}
    \caption{}
    \label{fig:setup_b}
\end{subfigure}
\hspace{0.5em}
\begin{subfigure}{0.28\textwidth}
    \centering
    \includegraphics[width=\textwidth]{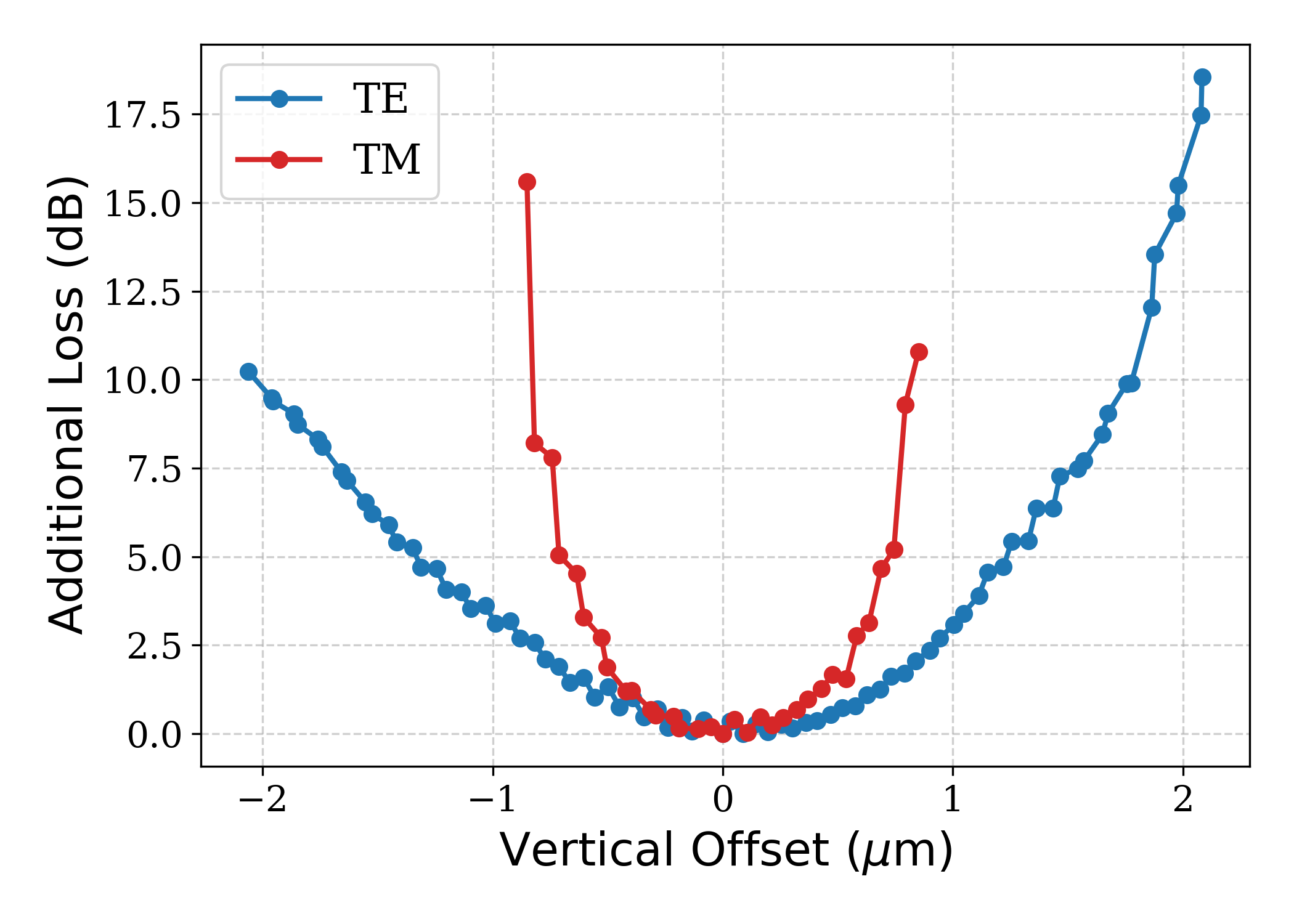}
    \caption{}
    \label{fig:setup_c}
\end{subfigure}
    \hspace{0.5em}
    \begin{subfigure}{0.28\textwidth}
        \centering
        \includegraphics[width=\textwidth]{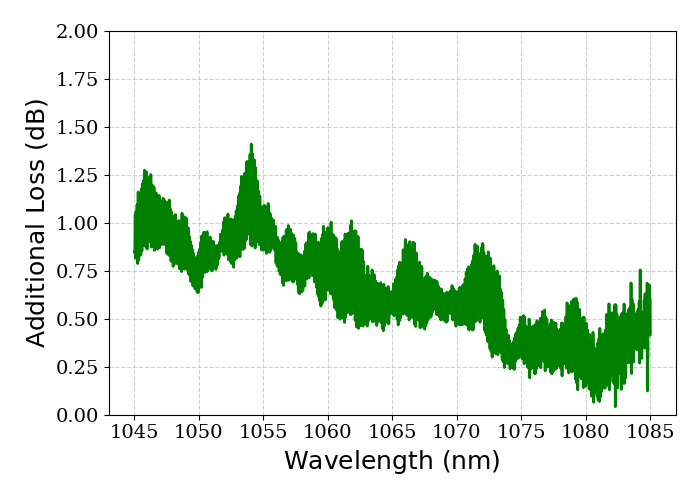}
        \caption{}
        \label{fig:setup_d}
    \end{subfigure}
    
    \caption{Experimental setup and measurement results of the fabricated photonic integrated circuit (PIC): (a) schematic of the overall measurement setup, (b) polarization-controlled additional loss as a function of lateral offset for TE-like and TM-like input polarization states, (c) polarization-controlled additional loss as a function of vertical offset for TE-like and TM-like input polarization states, and (d) additional loss as a function of wavelength.}
    \label{fig:setup}
\end{figure}

The completed devices were characterized using the setup illustrated in Fig.~\ref{fig:setup}. A Newport TLB~6700 tunable laser source (TLS) emitting at 1064 nm was launched as the input. The laser beam was first passed through a Thorlabs polarization controller to adjust the polarization state, and then coupled into the device under test (DUT) via an OZ Optics lensed fiber with a 2.5~$\mu$m mode field diameter (MFD). The output port used an identical 2.5~$\mu$m MFD lensed fiber to couple the transmitted light back into the collection fiber. The optical signal was then converted into an electrical signal using a Thorlabs photodiode for power measurement, or alternatively directed to a Yokogawa optical spectrum analyzer (OSA) to record the spectral response of the DUT. To evaluate the coupling loss per facet, a nanometer-resolution hexapod translation stage was used to precisely align the lensed fiber relative to the waveguide facet. The minimum insertion loss was determined when the photodiode reading reached its maximum. Since the optimized coupling parameters obtained from simulations were designed for the TE mode, the TM mode was expected to exhibit higher insertion loss. As such the polarization rotator was adjusted for maximum transmission.  Experimentally, a fiber-to-fiber insertion loss of 3.8 dB was measured. To analyze the contribution from waveguide propagation loss, we estimated an upper bound using a propagation loss of 0.8 dB/cm reported in our recent TFLN modulator work operating at 1064 nm\cite{de2026thin}. For the 2.5 mm long device used in this work, this value corresponds to only 0.20 dB total propagation loss, or approximately 0.10 dB per facet. Therefore, the propagation loss is much smaller than the measured fiber-to-fiber insertion loss, and the extracted coupling loss remains approximately 1.9 dB per facet. A 1064-nm mode-overlap analysis was performed to evaluate the contribution from the realistic layer stack, input taper-tip fabrication variation, and polished-facet angular deviation. For the 150-nm-wide and 150-nm-high taper input, the calculated overlap loss without the substrate effect is approximately 0.38 dB per facet. After including the realistic layer stack with a 1.5~$\mu$m top oxide, 4.7~$\mu$m BOX, and Si substrate, the calculated loss increases to approximately 0.72 dB per facet. A 15-nm reduction in both the input width and height further increases the calculated loss to approximately 1.85 dB per facet. In addition, the measured polished-facet angular deviation of approximately 4.5$^{\circ}$ can introduce about 0.51 dB per facet additional loss. These results indicate that the measured excess loss can be attributed to the combined effects of input cross-section fabrication variation and polished-facet angular deviation. In future devices, using a thicker BOX layer, such as a 10~$\mu$m BOX, together with improved polishing-angle control, should further reduce the coupling loss.
 Although this value is higher than the simulated prediction, to the best of our knowledge, this represents the lowest experimental demonstration of mode coupling at 1064 nm using a bilayer taper structure.  A systematic literature survey of the progress to date on end facet couplers for TFLN waveguides shows that, at 1550 nm, the lowest coupling loss achieved using bilayer inverse tapers can be as low as 0.5 dB per facet\cite{hu2021high}. Likewise, Ze et al. demonstrated an inverse taper at 1310 nm with a coupling loss of 1 dB per facet \cite{guo2024polarization}, Zhang et al. demonstrated an inverse taper at 1064 nm with a coupling loss of 3 dB per facet \cite{zhang2026cltwe} and, in the visible band at 775 nm, Xiao et al.\cite{liu2022ultra} reported a bilayer inverse taper with a coupling loss of 3 dB per facet. The measured coupling loss of 1.9 dB is consistent with the trend in the literature that the coupling loss improves with increasing wavelength, as shown in Fig. \ref{fig:wavelength_coupling}. We attribute this behavior to the fact that optical mode expands more readily at longer wavelengths, allowing better matching with the incident fiber mode. The figure also presents inverse taper data from an additional fabricated sample that uses the parameters reported in this paper at 1550 nm, demonstrating that the fabrication performance at this wavelength reaches or approaches the best results reported in the literature.

The alignment tolerance of the device was measured in the lateral and vertical directions with the input polarization fixed, as shown in Fig.~\ref{fig:setup_b} and Fig.~\ref{fig:setup_c}. The measured results indicate that the lateral alignment is less sensitive than the vertical alignment. For the TE polarization state, the additional loss remains below 1~dB over the measured range, whereas the vertical 1-dB tolerance is approximately 1.0~$\mu$m. For the TM polarization state, the vertical alignment tolerance is narrower, with a 1-dB width of approximately 0.74~$\mu$m. This behavior is in a good agreement with the polarization-dependent taper tip modal distribution. Because the taper tip cross section geometry supports different mode profiles for TE and TM polarizations, the fiber to taper mode overlap can be broader in the lateral direction than in the vertical direction. As a result, vertical alignment offset quickly reduces the overlap with the lensed fiber mode. These results indicate that the measured alignment tolerance depends on both the input polarization state and the end facet modal distribution.
The optical bandwidth of the device was also characterized, as shown in Fig.~\ref{fig:setup_d}. Within the wavelength window of 1055--1085 nm, the additional insertion loss remained below 1 dB, demonstrating that the device maintains a broad optical bandwidth and stable performance across this spectral range.Overall, the alignment tolerance and bandwidth response indicate that the device is promising for future robust optical packaging.

\section{Conclusion}

The work presents the design, fabrication, and characterization of a TFLN bilayer inverse taper device to support fiber to chip coupling at 1064 nm. The simulation results showed, with optimized parameters, the designed taper structure achieved a coupling loss of 0.48 dB per facet with a 2.5~$\mu$m MFD lensed fiber. The devices were fabricated and experimentally characterized and demonstrated a coupling loss of 1.9 dB per facet. The fabricated device exhibited a 1 dB optical bandwidth spanning 1055--1085 nm and reasonable tolerance to misalignment. Further work is required to improve fabrication tolerances and reduce the gap between the simulated and measured coupling performance; however, the demonstrated performance aligns well with demonstrated trends in literature for coupling loss as a function of wavelength. This development of an efficient fiber-to-chip edge coupler for TFLN at 1064 nm paves the way for future development of RF Photonic systems that can leverage higher performance devices at these wavelengths.

\begin{backmatter}

\bmsection{Acknowledgment}
The authors gratefully acknowledge the helpful discussions with the staff of the University of Delaware NanoFabrication Facility.

\bmsection{Disclosures}
The authors declare no conflicts of interest.

\bmsection{Data availability}
Data underlying the results presented herein are not publicly available at this time but may be obtained from the authors upon reasonable request.

\end{backmatter}

\bibliography{reference}

\end{document}